\documentstyle[twocolumn,prl,aps]{revtex}  
\begin{document}
\draft
\title{Transition Matrix Monte Carlo Reweighting and Dynamics}

\author{Jian-Sheng Wang$^1$, Tien Kiat Tay$^1$, and Robert H. Swendsen$^2$}

\address{$^1$Department of Computational Science,
National University of Singapore,
Singapore 119260, Republic of Singapore}

\address{$^2$Department of Physics,
Carnegie Mellon University,
Pittsburgh, PA 15213}

\date{10 September 1998}

\maketitle

\begin{abstract}

We study an induced dynamics in the space of energy of
single-spin-flip Monte Carlo algorithm.  The method gives an efficient
reweighting technique. This dynamics is shown to have relaxation times
proportional to the specific heat.  Thus, it is plausible for a
logarithmic factor in the correlation time of the standard 2D Ising
local dynamics.

\end{abstract}


The investigation of new Monte Carlo algorithms has been actively
pursued due to their importance for efficient computer simulation.
Cluster algorithms
\cite{Swendsen-Wang-PRL96,Swendsen-Wang-Ferrenberg,Wolff},
generalized ensemble methods \cite{Berg}, together with advanced
techniques of analysis, such as the histogram methods
\cite{Ferrenberg-Swendsen,M-histo,Oliveira,Oliveira-exact}, greatly
enhanced our ability to do efficient computer simulations.

Recently, Swendsen and Li \cite{Swendsen-TMC} proposed a novel
algorithm named transition matrix Monte Carlo (TMMC).  The method
first determines a transition matrix $W(E|E')$, to be defined below,
by some sampling technique.  The invariant state of this matrix gives
the canonical probability distribution of energy $P_{eq}(E)
\propto n(E) \exp(-E/k_BT)$, where $n(E)$ is the density of states.
This method has connections with the broad histogram method 
\cite{Oliveira}, the projection dynamics \cite{Novotny}, 
and the canonical transition probability method \cite{Fitzgerald}, but 
its starting point and considerations are quite different. 

There are three important aspects in this approach: (1) The
computation of the matrix can be efficiently implemented with the
$N$-fold way \cite{BKL}; (2) the method gives a new reweighting
technique, efficient and simple in comparison with multiple histogram
method \cite{M-histo}; (3) the artificial dynamics is of considerable
theoretical interest---it could imply a logarithmic correction in the
two-dimensional Ising model critical relaxation, indirectly supporting
Domany's conjecture \cite{Domany}.

The rest of the paper is organized as follows: we first define the
transition matrix Monte Carlo (TMMC) dynamics.  We discuss the
reweighting method under TMMC.  We then describe the steps leading to
an exact result for $W(E|E')$ for the one-dimensional Ising model.  We
give a differential equation obeyed in the large-size limit and
indicate its consequence.  We give arguments for relaxation time in
one dimension and relate the relaxation time to specific heat.  We
present simulation data to verify the general results.

Let us consider a single-spin-flip Glauber dynamics \cite{Glauber} of the
Ising model which is described in continuous time as
\begin{eqnarray}
{\partial P(\sigma, t) \over \partial t} =  &&
\sum_{\{\sigma'\}} \Gamma(\sigma,\sigma') P(\sigma', t) \nonumber \\
 = && \sum_{i=1}^N \Bigl[ -w_i(\sigma_i) + w_i(-\sigma_i)F_i\Bigr] P(\sigma,t),
\label{Meq}
\end{eqnarray}
where $N$ is the total number of spins, and 
\begin{equation}
w_i(\sigma_i) = { 1\over 2}\biggl[ 1 - \sigma_i \tanh\Bigl( K\!\!\! 
                   \sum_{\hbox{nn of i}}\!\!\sigma_k\Bigr) \biggr],
\; K = {J \over k_BT},
\end{equation}
is the flip rate of site $i$, which depends on the spin value at the
site $i$ as well as the values of its nearest neighbor spins
$\sigma_k$.  And $F_i$ is a flip operator such that $F_i
P(\ldots,\sigma_i, \ldots) = P(\ldots,-\sigma_i, \ldots)$.  The
transition matrix Monte Carlo dynamics is {\sl defined}, based on the
above dynamics, by
\begin{equation}
 { \partial P(E,t) \over \partial t }  = \sum_{E'} W(E|E') P(E',t), \label{Teq}
\end{equation} 
where $P(E,t)$ is the probability of having energy $E$ at time $t$, and 
\begin{equation}
  W(E|E') = {1 \over n(E') } \sum_{H(\sigma) = E}\sum_{H(\sigma') = E'}
\!\!\! \Gamma(\sigma, \sigma').
\end{equation} 
Although Eq.~(\ref{Meq}) and (\ref{Teq}) give the same equilibrium
distribution of energy, they have totally different dynamics.

We note that the transition matrix is banded along diagonal.  The
matrix elements are not independent---the column sum is zero due to
the conservation of total probability, and the row sum satisfies
$\sum_{E} W(E'|E) P_{eq}(E) = 0$ due to the fact that the equilibrium
distribution is a stationary distribution.  The transition matrix also
satisfies detailed balance condition, $W(E|E')P_{eq}(E') =
W(E'|E)P_{eq}(E)$, inherited from the detailed balance in the original
dynamics of spins. The detailed balance conditions put strong
constraint on the matrix elements.  For example, for any three
energies with nonzero transition rates among them, we have
\begin{equation}
W(E|E'')W(E''|E')W(E'|E) = W(E|E')W(E'|E'')W(E''|E). 
\end{equation}

The matrix elements can be computed by Monte Carlo sampling as
follows. For a configuration $\sigma$, we consider the number
$N(\sigma, \Delta E)$ of cases that energy is changed by $\Delta E$,
for the $N$ possible single-spin flips.  Then for $\Delta E \neq 0$,
\begin{equation}
  W(E+\Delta E|E) = w(\Delta E) 
\bigl\langle N(\sigma, \Delta E) \bigr\rangle_{H(\sigma)=E},
\label{TNeq}
\end{equation} 
where the average is over all configurations having energy $E$, and
$w(\Delta E) = {1\over 2} \bigl( 1 - \tanh(\Delta E/(2k_BT)\bigr)$ for
the Glauber dynamics.  Since the quantity in the angular brackets of
Eq.~(\ref{TNeq}) is independent of temperature (or flip rates), once
it is determined, we can use it at any temperature (or with any flip
rates) for $W(E|E')$ and consequently the equilibrium distribution at
any temperature.

The microcanonical averages can be computed in any ensembles which
have the property that equal energy states have equal probability.  We
used canonical simulations at a selection of temperatures, so that the
total histogram $H(E)$ is approximately flat.  Of course, the total
$H(E)$ obtained by adding results at different $T$ is not meaningful.
However, it is perfectly correct to add statistics to $\bigl\langle
N(\sigma, \Delta E) \bigr\rangle_{E}$ from equilibrium simulations at
different temperatures.

The single histogram method \cite{Ferrenberg-Swendsen} uses $H(E)$ in a
canonical simulation as an estimate to the equilibrium energy
distribution $P(E)$.  Multiple histogram method attempts to overcome
the problem of narrow window in the distribution
\cite{M-histo}.  In the present method, adding simulations at
different $T$ is handled rather naturally.  Such a ``multiple
histogram'' TMMC is extremely simple and effective.  In connection
with the broad histogram method \cite{Oliveira}, we note that when $T$
is set to $\infty$, the detailed balance conditions reduce to the
broad histogram equations \cite{Oliveira-exact}.  Unfortunately, the
dynamics proposed in Ref. \cite{Oliveira} is incorrect
\cite{Hansmann}.

\begin{figure}
\epsfxsize=\hsize\epsffile{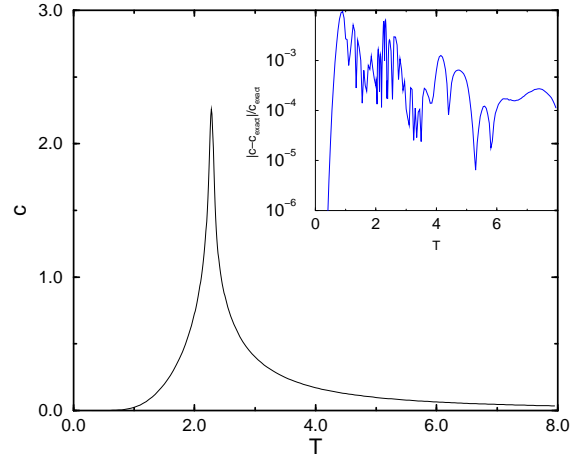}
\caption[fig0]{The specific heat of the two-dimensional Ising model on a 
$64^2$ lattice by the transition matrix Monte Carlo reweighting
method.  The insert shows the relative error with respect to the exact
result (obtained numerically based on \cite{exact}).  The simulations
are done at 25 temperatures, each with $10^6$ Monte Carlo steps with a
single-spin-flip dynamics.  }
\label{fig-1}
\end{figure}

There are a number of different ways of determining the canonical
distributions from $W(E|E')$.  We can solve the linear equation, $\sum_{E'}
W(E|E') P(E') = 0$, or use the detailed balance condition $W(E|E')
P(E') = W(E'|E) P(E)$.  The latter seems numerically more stable---the
equations give us a set of recursion relations for $P(E)$.  In
Fig.~\ref{fig-1} we present a calculation of the Ising model heat
capacity on a 64 by 64 lattice with 25 simulations at selected
temperatures.  The errors are small for the whole temperature range.

The one-dimensional dynamics of spins, Eq.~(\ref{Meq}), can be solved
analytically \cite{Glauber}.  In particular, the full set of
eigenvalues of $\Gamma$ is known \cite{Felderhof}.  We show that the
matrix $W(E|E')$ can be obtained in closed form in one dimension.  To
begin with, the density of states is $n(E) = 2 C^L_{2k}$, where $k=0,
1, 2, \ldots, \lfloor L/2 \rfloor$, assuming periodic boundary
condition.  The number of possible states with energy $E/J= -L + 4k$
is twice the number of ways (due to overall up-down symmetry) of
putting $2k$ unsatisfied bonds among $L$ possible positions.

The summation over spin states with energy $E$ and $E'$ can be
expressed in terms of a one-dimensional lattice gas problem.  In one
dimension, the flip rate can be written as $w_i(\sigma_i) = (1/2)
\bigl[ 1 - \gamma
\sigma_i(\sigma_{i-1} + \sigma_{i+1})/2\bigr]$
where $\gamma = \tanh 2K$.  Notice that for each final configuration
$\sigma$ associated with $E$, contributions to the transition from
$E^+=E+4J$ to $E$ come only from each pair of nearest neighbors
satisfied bonds in configuration $\sigma$.  Thus we can write
\begin{equation}
 \Delta^+_k = {n(E^+) W(E| E^+)\over 1 + \gamma}
 = \!\!\!\sum_{\sum n_i = L-2k}\!\!\! \sum_i n_i n_{i+1},
\end{equation}
where $n_i = 0$ and 1 for unsatisfied and satisfied bond,
respectively.  Similar expressions can be written down associated with
the transition from $E^-= E - 4J$ to $E$.  The restricted sums are
obtained through a partition function in a field as a generating
function. We find $\Delta^+_k = L C^{L-2}_{2k}$ and $\Delta^-_k = L
C^{L-2}_{L-2k-2}$ and the transition matrix elements are thus
\begin{eqnarray}
    W_{k, k+1} = && { (k+1)(2k+1) \over L - 1} (1+\gamma), \\
    W_{k+1,k} = && { (L-2k)(L-2k-1) \over 2 (L-1) } (1-\gamma).
\end{eqnarray}
The diagonal term is computed from the relation
\begin{equation}
    W_{k-1,k} + W_{k,k} + W_{k+1,k} = 0,
\end{equation}
and the rest of the elements $W_{k,k'} = 0$ if $|k-k'| > 1$.

The matrix can not be diagonalized analytically in general.
Nevertheless, at zero temperature, $\gamma = 1$, the eigenvalues can
be obtained explicitly as $\lambda_k = -2(k+1)(2k+1)/(L-1)$, which
give us relaxation times as $-1/\lambda_k$, with the longest
relaxation time $\tau = (L-1)/2$.

The TMMC for large system follows a simple and interesting dynamics.
It can be shown rigorously, with the method of $\Omega$-expansion for
master equation \cite{VanKampen}, that in the large-size scaling
limit, the process is described by the equation
\begin{equation}
   {\partial P(x,t') \over \partial t'} = 
{ \partial \over \partial x } \left( {\partial P(x,t') \over \partial x }
+ x P(x,t') \right), \label{diffusionEq}
\end{equation}
where $t'$ and $x$ are properly scaled time and energy deviation from
equilibrium.    
\begin{equation} 
  x = { E - u_0 N \over ( N c' )^{1/2} },  \quad u_0 N = \bar E,
\end{equation}
and $t' = b t$ with
\begin{equation}
  b = \lim_{N\to\infty} {1\over 2 c' N} \sum_{E} W(\bar E|E) (E-\bar E)^2,
\label{beq}
\end{equation}
where $u_0$ is the average energy per spin and $c'=k_BT^2c$ is the
reduced specific heat per spin.  The equation is obtained by replacing
$E$ by $x$ and expanding all the terms in small parameter
$1/\sqrt{N}$.  Details will be presented elsewhere.

The continuum limit equation describes a constrained random walk.
There are two competing effects in the current $j = - \partial
P/\partial x - x P$; while the first term is the usual diffusion, the
second term keeps the walker at the origin.  Equilibrium is obtained
when $j=0$, giving the well-known Gaussian distribution $P_{eq}(x)
\propto e^{-x^2/2}$.

Equation (\ref{diffusionEq}) can be transformed into a one-dimensional
quantum harmonic oscillator equation with the change of variable
$P(x,t') = e^{-x^2/4} \phi(x,t')$.  The relaxational spectrum is
discrete and equally spaced (in $1/\tau$).  In particular, the
relaxation modes are $e^{-nt'- x^2/2} H_n(x/\sqrt{2})$, where $H_n$ is
Hermite polynomials.  Translating back to the original time $t$, the
relaxation times are $1/(nb) \propto c$.  Applying the general result,
Eq.~(\ref{beq}), to the one-dimensional Ising model, we have
\begin{equation}
\tau_n = {1 \over 2 n } \cosh 2K, \quad n = 0, 1, 2, \cdots.
\end{equation}

It is instructive to have an intuitive picture of the asymptotic
dynamics, which we argue as follows.

The TMMC is equivalent to the following steps (Algorithm A).
\begin{enumerate}
\item  Do perfect microcanonical simulation, i.e., pick a state
at random among the degenerate states of current energy $E$.

\item  Do one canonical Monte Carlo move, chosen a site at random.
\end{enumerate}
Repeat step (1) and (2) $N$ times, where $N$ is the number of spins in
the system.  This is one TMMC step (sweep).

Consider the dynamics at very low temperatures.  Then only two
energies $E_0$ (ground state) and $E_1=E_0+4J$ (first excited states)
dominate.  Consider $K$ such that correlation length ($\xi \propto
\exp(2K)$) is about the size of the system.  $ P(E_0) \propto
\exp(-E_0/k_BT)$, $P(E_1) \propto L^2 \exp\bigl((-E_0
-4J)/k_BT\bigr)$, and $P(E_1)/P(E_0) \approx L^2 \exp(-4K)
\approx 1$.
 
We now consider the time scales (from the point of view of TMMC) that
a transition is made $E_0 \to E_1$, and $E_1 \to E_0$.

Let's assume that the system is in its ground state $E_0$.  Step (1)
actually does nothing; in step (2), a kink pair (unsatisfied bonds) is
excited with probability $\exp(-4K)$.  Thus, we need $\exp(4K)$ moves
to produce a kink pair.  So the time scale in units of TMMC steps is $
\tau \propto \exp(4K)/L \propto L$.  We have used the fact that
correlation length in one dimension is proportional to $\exp(2K)$ and
that correlation length is comparable to system size $L$.  The reverse
process has a similar scale. This argument is consistent with the
exact calculation.
 
A more general argument \cite{Swendsen} relating the relaxation time
to the specific heat can be given, as follows: the variance of the
energy distribution $P(E)$ is related to the specific heat as $\delta
E^2 = c\, N k_BT^2$.  Transition matrix Monte Carlo moves are random
walks in energy confined in a region of $\delta E$.  Thus the typical
time for energy varying over $\delta E$ is proportional to the square
of distance in energy.  Therefore, time is proportional to $\delta E^2
$, when we measure in steps of moves.  To get $\tau$ in sweeps, we
must divide by $N$.  Let $a$ be the typical time for one step of walk
in energy, the equation should be
\begin{equation}
\tau \propto a (\delta E^2/N) =  a\, c\,k_BT^2 = a\, c',
\end{equation}
We can give $a$ a precise meaning as $1/(c'b)$, where $b$ is given in
Eq.~(\ref{beq}).  In one dimension, the unit time $a$ diverges.
However, this does not happen in dimensions $d>1$ as $T_c>0$, and $a$
will be some finite value asymptotically independent of $N$.

\begin{figure}
\epsfxsize=\hsize\epsffile{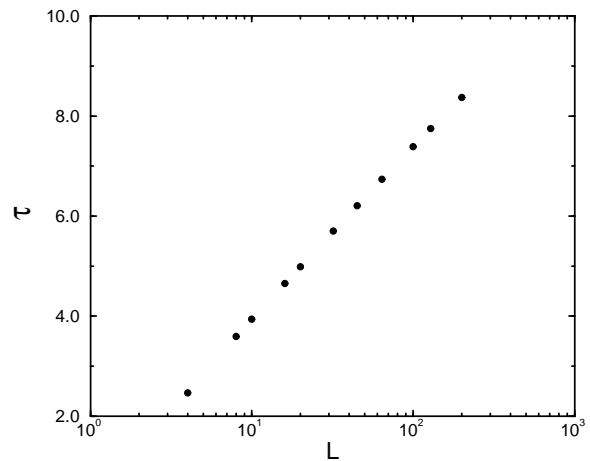}
\caption{Relaxation time for two-dimensional Ising model at $T_c$
of the transition matrix Monte Carlo dynamics computed from the inverse
of the greatest non-zero eigenvalues of matrix $W(E|E')$.  The
matrices are obtained by a Wolff Monte Carlo cluster algorithm with
$10^7$ to $10^9$ cluster flips.}
\label{fig-2}
\end{figure}

We check the above results in two dimensions by extensive Monte Carlo
simulation using two different methods: (1) the
microcanonical/canonical algorithm A described above, by computing the
time correlation functions; and (2) by direct computation of $W(E|E')$
with a Wolff cluster algorithm \cite{Wolff}.  The eigenvalues are
computed with standard package \cite{Lapack}.  While the first method
is restricted to very small sizes, the second method can apply to
large systems. In Fig.~\ref{fig-2} we plot only the inverse
eigenvalues.  Very good confirmation of the $\log L$ behavior for the
relaxation time is observed.

This result has serious implication.  In the standard single-spin-flip
dynamics, if we add between canonical Monte Carlo moves microcanonical
moves, the resulting dynamics is TMMC and still has a residue slowing
down $\tau \sim \log L$ at $T_c$.  It is thus difficult to imagine how
this logarithmic factor can be canceled exactly in the original
dynamics.  In fact, such logarithmic factor is conjectured
\cite{Domany}, while many numerical computations \cite{Wang-z}
did not seem to find it explicitly.  We think that the Domany conjecture
is still an open question.

\begin{figure}
\epsfxsize=\hsize\epsffile{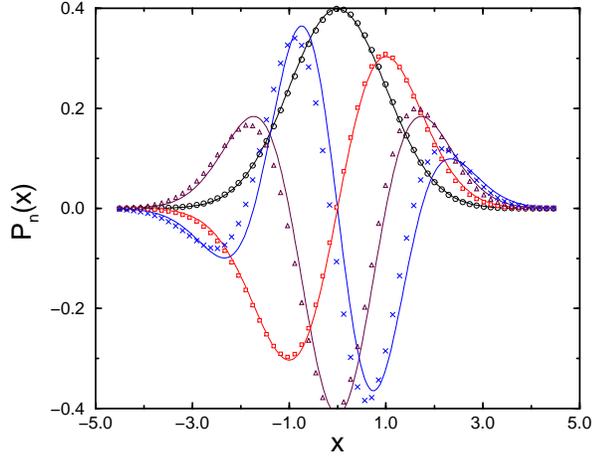}
\caption[fig3]{The first four eigenmodes of the transition matrix
Monte Carlo dynamics.  The continuous curves are analytic results
$\propto e^{-x^2/2} H_n(x/\sqrt{2})$.  Symbols are from exact
diagonalizations of matrix $W(E|E')$ for a three-dimensional Ising
model on a $16^3$ system at $k_BT/J = 6.0$.  For clarity, only every
fifth data points are plotted.  }
\label{fig-3}
\end{figure}

One of the prediction of Eq.~(\ref{diffusionEq}) is that the
eigenvalues of $W(E|E')$ in the large-size limit are equally spaced,
this is indeed the case for large systems.  The eigenfunctions
associated with these eigen modes are compared with numerical results.
In Fig.~\ref{fig-3}, we plot the analytic results (curves) together
with numerical values (symbols) from exact diagonalization of matrix
$W(E|E')$ for a three-dimensional Ising model of $16^3$ at $k_BT/J =
6.0$.  There are no adjustable parameters in the comparison except an
overall normalization.

In conclusion, the transition matrix Monte Carlo shows a novel
dynamical behavior.  We find an unusual critical slowing down in one
dimension.  For two dimensions and higher, we conclude that
correlation time is proportional to the specific heat.  This is in
contrast with cluster algorithms where Li and Sokal \cite{Li-Sokal}
showed that the specific heat is only a lower bound to the correlation
time.  While the TMMC artificial dynamics is of theoretical interest,
the reweighting technique is very useful in practice.

J.~S.~W thanks Lei-Han Tang for an important suggestion on large-size
limit.  He also thanks Hong Guo and Donghui Zhang for discussions.
This work is supported in part by Academic Research Grant
No.~RP950601.

\input epsf.tex

\end{document}